\documentclass{article}

\usepackage{amsmath}
\usepackage{commath}
\usepackage{graphics}
\usepackage{amsfonts}
\newcommand{\CC}{{\mathbb C}}
\newcommand{\RR}{{\mathbb R}}
\newcommand{\ZZ}{{\mathbb Z}}
\newcommand{\NN}{{\mathbb N}}

\def\squareforqed{\hbox{\rlap{$\sqcap$}$\sqcup$}}
\def\qed{\ifmmode\else\unskip\quad\fi\squareforqed}

\newtheorem{theorem}{Theorem}[section]
\newtheorem{lemma}[theorem]{Lemma}

\newtheorem{corollary}[theorem]{Corollary}
\newtheorem{conjecture}[theorem]{Conjecture}

\newenvironment{proof}[1][Proof]{\begin{trivlist}
\item[\hskip \labelsep {\bfseries #1}]}{\end{trivlist}}

\newenvironment{remark}[1][Remark]{\begin{trivlist}
\item[\hskip \labelsep {\bfseries #1}]}{\end{trivlist}}

\begin{document}

\title{Integrable deformations of CFTs and the discrete Hirota equations}

\author{Werner Nahm\thanks{wnahm@stp.dias.ie} \and Sin\'ead Keegan\thanks{skeegan@stp.dias.ie}}

\date{{\small Dublin Institute for Advanced Studies, \today}}



\maketitle

\begin{abstract}
We solve the discrete Hirota equations (Kirillov-Reshetikhin Q-systems) for $A_r$, and their analogue for $D_r$, for the cases where the second
variable ranges over either a finite set or over all integers. Until now only special solutions were known. We find all solutions for which no component vanishes, as required in the known applications. As an introduction we present the known solution where the second variable ranges over the natural numbers.
\end{abstract}

\section{Introduction}

\label{intro}
Many conformally invariant quantum field theories in two dimensions (CFTs) can be described as limits of integrable massive quantum field
theories. In such a situation, certain properties of the CFT can be understood from the perspective of the massive theory, in particular its
spectrum. Suppose that the massive theory has $K$ species of fermionic particles labeled by $a=1,\ldots,K$, and a diagonal scattering matrix
$S_{ab}=\exp(i\Phi_{ab}(\theta^a-\theta^b))$ for particles of types $a,b$ with rapidities $\theta^a,\theta^b$. Then the expected spectrum of
the limiting CFT can be described as follows. Let $\Delta\Phi$ be the $K\times K$ matrix with entries
$\Delta\Phi_{ab}=\Phi_{ab}(+\infty)-\Phi_{ab}(-\infty)$ and put $ A=I-\Delta\Phi/2\pi$, where $I$ denotes the identity matrix. Then for each
conformal block of the CFT the partition function has the form
\begin{equation}\label{fp}
Z_L = \sum_{N\in\NN^K}\frac{q^{Q(N)}}{(q)_{N_1}\cdots (q)_{N_K}},
\end{equation}
where we use the standard notation $(q)_n=(1-q)\cdots (1-q^n)$ for the $q$-factorial, and where
\begin{equation}\label{QN}
Q(N)=\frac{1}{2}\sum_{a,b=1}^K N_a A_{ab} N_b + L(N).
\end{equation}
$L$ is an affine linear function which depends on the conformal block. One needs non-negative $A$ for equation~(\ref{fp}) to make sense. We
shall assume that $A$ is positive. Heuristically, the occurence of such a partition function can be understood as follows. Consider a system
with spatial periodicity $R$. For $N_a$ free fermions of types $a=1,\ldots,K$, with antiperiodic boundary conditions, the contribution to the
partition function is
\begin{equation}
\frac{q^{\sum N_a^2/2}}{(q)_{N_1}\cdots (q)_{N_K}}.
\end{equation}
This yields equation~(\ref{fp}) when $A=I$ and $L=0$. To see how this contribution changes when $\Phi_{ab}$ does not vanish, consider the
Bethe ansatz equations for $N_a$ particles of type $a$ with rapidities $\theta^a_i$, where $i=1,\ldots,N_a$, and $a=1,\ldots, K$. They take
the form
\begin{equation}\label{bethe}
m_a R \sinh(\theta^a_i)+\sum_{b=1}^K {\sum_{j=1}^{N_b}}' \Phi_{ab}(\theta^a_i-\theta^b_j)=2\pi n^a_i,
\end{equation}
where $m_a$ is the mass of a particle of type $a$, and the $n^a_i$ are positive integers (or half integers in the antiperiodic case). The
prime on the summation sign means that for $b=a$ the term with $j=i$ is to be omitted. For small $R$, particles moving in opposite directions
should decouple, so that one can put $\Phi_{ab}(+\infty)=0$ in the description of right movers. The quantisation of the total momentum of the
system is not affected by the interaction. For a system of two particles of types $a,b$ this means that
$\Phi_{ab}(\theta) + \Phi_{ba}(-\theta)$ must be independent of $\theta$, in agreement with standard constraints on the scattering matrix.
This implies that
$\Phi_{ab}(\theta) + \Phi_{ba}(-\theta) = -\Delta\Phi_{ab}$,
since this is true for $\theta = \infty$.

Now let all $\theta^a_i$ be positive. Then equation~(\ref{bethe}) implies
\begin{equation}
 \sum_{a=1}^K\sum_{i=1}^{N_a} m_a R \sinh(\theta^a_i) =-\pi\sum_{a,b=1}^K N_a \Delta\Phi_{ab} N_b + 2\pi\sum_{a=1}^K\sum_{i=1}^{N_a}n^a_i.
\end{equation}
In the conformally invariant limit, rapidities are large and the difference between $\sinh(\theta)$ and $\cosh(\theta)$ can be neglected.
Thus the previous equation yields the energy of the system. This differs from the free fermion case by an additive term, in agreement with
equation~(\ref{QN}). The linear part, $L(N)$, depends on the number of particles with zero rapidity and on the zero point energy, which needs
a different treatment. The argument just given is heuristic, but in some cases exact formulae for the spectrum of the Hamiltonian have been
obtained which agree with the preceding argument in the conformal limit.

Partition functions of conformal theories must be modular. This implies that they can be written as a sum over expressions of the form
$\tilde q^{h-c/24}$, where $\log q \log\tilde q= 4\pi^2$, $h$ is a conformal dimension of the theory and $c$ its central charge. The smallest
conformal dimension yields the exponent $-c_{eff}/24$. Evaluating $Z_L$ close to $q=1$ yields
\begin{equation}
\frac{\pi^2}{6}\, c_{\mathit eff}=\sum_{i=1}^K L(1-x_i),
\end{equation}
where $L$ is Roger's dilogarithm and $x=(x_1,\ldots,x_K)$ is the unique solution of the system of equations
\begin{equation}\label{1-x}
 \sum_{j=1}^K A_{ij}\log(x_j) = \log(1-x_i),
\end{equation}
with $i=1,\ldots,K$ and $(x_1,\ldots,x_K)\in (0,1)^K$.

According to an idea of Zagier the existence of a unique solution can be proved as follows. Consider the map $\phi:(0,1)^K\rightarrow \RR^K$
given by
\begin{equation}
\phi_i(x)=\sum_{j=1}^K A_{ij}\log x_j -\log(1-x_i).
\end{equation}
It is sufficient to show that $\phi$ is a homomorphism. Locally this is true since the Jacobian matrix with components
$J_{ij}=\partial\phi_i/\partial x_j$ is nowhere degenerate. Indeed the product matrix $J\mathrm{diag}(x_j)$ is everywhere positive. Moreover
one easily sees that the inverse image under $\phi$ of every compact domain in $\RR^K$ is compact as follows. Let $C=K(\log 2)^2$ be the
maximum of $\sum_{j=1}^K \log x_i\log(1-x_i)$. Then $\phi_i(x)>a_i$ implies
\begin{equation}
 \sum_{i,j=1}^K \log x_i A_{ij}\log x_j - \sum_{i=1}^K a_i\log x_i < -C,
\end{equation}
such that $\left(\log x_1,\ldots,\log x_K\right)$ lies in an ellipsoid. Replacing $x_i$ by $1-x_i$, and $A$ by $A^{-1}$, one sees that the
same is true for $\left(\log (1-x_1),\ldots,\log (1-x_K)\right)$. Let $x^{(1)}, x^{(2)},\ldots$ be any sequence in $(0,1)^K$. If
$\phi\left(x^{(k)}\right)$ converges in $\mathbb{R}^K$, the $x^{(k)}$ lie in a compact set and must have an accumulation point. This implies
that $\phi\left(x^{(k)}\right)$ converges to an element in the image of $\phi$. Since the image of $\phi$ is both open and closed, it must be
equal to $\RR^K$. It follows, by the monodromy theorem, that $\phi$ is a homomorphism.

To calculate non-leading terms in the modular transformation of $Z_L$ one needs complex solutions of equation~(\ref{1-x}), where for each $i$
one of the possible values of $\log x_i$ and $\log (1-x_i)$ has to be chosen. The corresponding values of a suitably defined version of the
Rogers dilogarithm yield the exponents $h-c/24$, multiplied by $\pi^2/6$. The main purpose of the present paper is the solution of
equation~(\ref{1-x}) in some special cases. In this context it will be necessary to solve certain infinite systems of equations of
independent interest.

Despite the occurence of logarithms, equation~(\ref{1-x}) is essentially algebraic since exponentiation yields
\begin{equation}\label{alg}
 \prod_{j=1}^K x_j^{A_{ij}} = 1-x_i.
\end{equation}
Omitting indices, these equations will be written as $x^A=1-x$. Since logarithms must be taken, one is only interested in solutions for which
all $x_i$ are different from $0$ and $1$. These solutions yield torsion elements of the algebraic K-theory group~\cite{Nahm,Neumann}. This
property can be used to restrict the possible candidates for the matrix $A$. In particular, $K=1$ allows only the three cases $A=2,1,1/2$,
see~\cite{Zagier2}. More generally, one may have
\begin{equation}
 A={\cal C}(X)\otimes {\cal C}(Y)^{-1},
\end{equation}
where $X$ and $Y$ are Dynkin diagrams of type ADET and $\mathcal{C}$ denotes the corresponding Cartan matrices. Here $K$ is the product of
the ranks of $X$ and $Y$, so for $K=1$ one needs $\mathcal{C}(A_1)=2$ and/or $\mathcal{C}(T_1)=1$, in agreement with $A=2,1,1/2$. We denote
the set of vertices of $X$ and $Y$ by $V(X)$ and $V(Y)$ respectively, and write $V(X,Y)=V(X)\times V(Y)$, so that the components of $x$ have
the form $x_{im}$ with $(i,m)\in V(X,Y)$. We will also consider more general auxiliary diagrams with an infinite number of vertices, but all
will be simply laced, with a finite number of vertices adjacent to each given one. As usual, we put $V(A_r)=\{1,\ldots,r\}$, with
adjacencies between successive integers. We also introduce diagrams $Y=A_\NN$ and $Y=A_\ZZ$ with $V(Y)=\NN,\ZZ$, respectively, and
adjacencies between successive integers.

With
\begin{equation}
\log \left(x_{im}\right) = \sum_{n\in V(Y)} \mathcal{C}(Y)_{mn}\log\left(z_{in}\right),
\end{equation}
one obtains
\begin{equation}\label{X,Y}
z^{(2-{\cal C}(X))\otimes I}+z^{I\otimes (2-{\cal C}(Y))}=z^2,
\end{equation}
where indices and identity matrices have been suppressed in the notation. Given diagrams $X$ and $Y$, we shall refer to the corresponding
equations~(\ref{X,Y}) as $(X,Y)$ equations, with one component equation for each pair $(i,n)\in V(X,Y)$. Solutions are considered as elements
$z\in\CC^{V(X,Y)}$. Since Cartan matrices are invertible, $\log z_{in}$ must be finite for every $(i,n)\in V(X,Y)$, such that no component of
$z$ vanishes. The set of such solutions $z\in(\CC^*)^{V(X,Y)}$ will be denoted $S^*(X,Y)$. A first result which shows that the $(X,Y)$
equations simplify equation~(\ref{1-x}) is the following.

\begin{theorem}
When $X$ and $Y$ are Dynkin diagrams of type ADET, the components of any solution $z$ of the $(X,Y)$ equations are algebraic integers.
\end{theorem}

\begin{proof}
It suffices to show that the $\ZZ$-span of arbitrary monomials in the $z_{in}$ has finite rank. Let the height of a monomial
$\prod_{in}z_{in}^{N_{in}}$ be given by $\sum_{in}f_X^if_Y^n N_{in}$, where $f_X,f_Y$ are the Frobenius eigenvectors of $\mathcal{C}(X)$,
$\mathcal{C}(Y)$. When $z_{in}^2$ is replaced by the left hand side of the $(i,n)$ component of the $(X,Y)$ equations, the monomial is
represented as the sum of two monomials of lower height. The height is lowered by at least the Frobenius eigenvalue of the Cartan matrices.
After a finite number of such substitutions one is thus left with a sum of monomials for which all exponents are zero or one. There is only a
finite number of such monomials. \qed
\end{proof}

In the following we will consider only diagrams $Y$ of type $A$. To solve the $(X,A_n)$ equations it is helpful to first consider $(X,A_\NN)$.
The corresponding system of equations has been studied extensively, for example in~\cite{KdiF,H,N}, and is closely connected to the
representation theory of Yangians~\cite{KR}. Let $X$ have rank $r$. Consider the Yangian $Y(X)$, which contains the enveloping algebra $U(X)$
as a sub-Hopf-algebra. The irreducible finite-dimensional representations of $Y(X)$ have highest weights $\lambda=\sum_{i=1}^r n_i\lambda_i$,
which are obtained by restricting them to representations of $X$. Here  the $\lambda_i$ are the fundamental weights of $X$ and the $n_i$ are
non-negative integers. The representations with highest weight $\lambda$ are characterised by $r$ monic polynomials of degrees
$n_1,\ldots,n_r$. For $\lambda_0\in\RR$ there is an outer automorphism of $Y(X)$ which preserves $U(X)$ and acts on the polynomials by a shift
$\lambda\mapsto\lambda+\lambda_0$. For $\lambda=n\lambda_i$ the irreducible representations are characterised by a single polynomial of order
$n$. There is a basic representation given by the polynomial $\prod_{i=1}^n (\lambda-i)$ (up to shifts of $\lambda$). Let $\chi_{ni}$ be the
corresponding character of $X$. These characters decompose into sums of irreducible characters which may be quite complicated, see
e.g.~\cite{Kleber}. Note that $\chi_{0i}(g)=1$ for any $g\in Lie(X)$. The $(A_r,A_\NN)$ equations were solved by Kirillov~\cite{Kirillov},
and the $(D_r,A_\NN)$ equations by Kirillov and Reshetikhin~\cite{KR}, though in the latter case no complete proof was published. More
precisely, they found the generic solution of these equations, which is all one needs in our context. For $X=A_r$ this solution can be
written as $z_{ni}=\chi_{in}(g)$, with arbitrary $g$, and they conjectured that this is true for $X$ of type $D$ or $E$, too. For a proof and
its history see~\cite{KdiF}, \cite{RK}.

The $(X,A_\NN)$ and $(X,A_\ZZ)$ equations can be considered as initial value problems. We formulate this property as follows. For
$Y=A_n,A_\NN$ let $R_1$ be the restriction map $R_1:\CC^{V(X,Y)}\mapsto \CC^{V(X)\times\{1\}}$ and for $Y=A_\ZZ$ let $R_{01}$ be the
restriction map $R_{01}:\CC^{V(X,Y)}\mapsto \CC^{V(X)\times\{0,1\}}$.

\begin{lemma}
For $m>0$ let $z,z'\in V(X)\times \{1,\ldots,m+1\}$ be solutions of the $V(X)\times \{1,\ldots,m\}$ components of the $(X,A_\NN)$ equations.
Let $R_1z=R_1z'$ and assume that no component of $z$ vanishes. Then $z'=z$. For $m\leq 1\leq m'$ let
$z,z'\in V(X)\times \{m-1,m,\ldots,m'+1\}$ be solutions of the $V(X)\times \{m,\ldots,m'\}$ components of the $(X,A_\ZZ)$ equations. Let
$R_{01}z=R_{01}z'$ and assume that no component of $z$ vanishes. Then $z'=z$.
\end{lemma}

\begin{proof}
First consider $Y=A_\ZZ$. Let $X(i)$ be the set of vertices of $X$ which are adjacent to a vertex $i$. Then the $(X,Y)$ equations imply
\begin{equation}\label{recX}
z_{i,m+1}=\frac{z_{im}^2-\prod_{k\in X(i)}z_{km}}{z_{i,m-1}},
\end{equation}
such that $z_{i,m+1}=z'_{i,m+1}$ by induction on $m$, and analogously $z_{i,m'-1}=z'_{i,m'-1}$ by induction on $-m'$. For $Y=A_\NN$ it is
sufficient to specialise the argument to $m=1$ and $z_{i0}=1$ for all $i\in V(X)$. \qed
\end{proof}

In the following we will need the set of solutions of the $(X,Y)$ equations which contains $S^*(X,Y)$ and all its limit points in
$\CC^{V(X,Y)}$. For convenience we use the language of algebraic geometry. Let ${\mathcal I}(X,Y)$ be the ideal of polynomials on
$\CC^{V(X,Y)}$ which vanish for all
$z\in S^*(X,Y)$. Let $S(X,Y)$ be the algebraic completion of $S^*(X,Y)$, i.e. the set of all $z$ which are zeros of all elements of
${\mathcal I}(X,Y)$. In general, the set of all solutions of the $(X,Y)$ equations has a countable infinity of other components, but these
are irrelevant for our purpose.

\begin{lemma}
Let $X$ be any connected unordered graph, with finite $X(i)$ for any $i\in X$. Then all components of $z\in S(X,A_\NN)$ are polynomials in
the components of $R_1z$, such that $S(X,Y)$ is an algebraic variety isomorphic to $\CC^{rk(X)}$ (or empty).
\end{lemma}

\begin{proof} We use induction on $m$. Assume that the $z_{in}$ are polynomials for $n\leq m$. Then equation~(\ref{recX})
yields $z_{i,m+1}$ as the quotient of two polynomials. To show that the quotient is a polynomial, it suffices to show that
$z_{im}^2-\prod_{k\in X(i)}$ vanishes whenever $z_{i,m-1}$ vanishes,
and at least at the same order. Now $z_{i,m-1}=0$ implies that $z_{i,m-2}^2=\prod_{k\in X(i)}z_{k,m-2}$ and
$z_{k,m-1}^2=z_{k,m-2}z_{k,m}$ for $k\in X(i)$. Thus $z_{i,m-2}^2(z_{im}^2-\prod_{k\in X(i)})=0$. By assumption $z_{i,m-2}\neq 0$ generically,
and one checks easily that the order of vanishing works out correctly. \qed
\end{proof}

This polynomial property was proved in \cite{FKII} using the Laurent phenomenon in cluster algebras, but our approach is much simpler.
In the following we only consider $X$ of type $A$ or $D$. We will need more detailed information
on the polynomial expressions for $z$, namely the explicit form of the polynomials given in \cite{Kirillov} and \cite{KNH}.

\section{The A Case}
\label{sec:1}

We first treat the case $X=A_r$. This yields a special case of the discrete Hirota equations, which has a well known explicit solutions.

\begin{theorem}\label{A1}
Let $x\in\CC^\ZZ$. For any  positive integer $m$, and for $i\in \ZZ$, let $M^m_i(x)$ be the $m\times m$ Toeplitz matrix with entries
\begin{equation*}
\left(M^m_i(x)\right)_{kl}=x_{l-k+i}.
\end{equation*}
Let $z_{im}=\det M^m_i(x)$. Then these $z_{im}$ satisfy the $(A_\ZZ,A_\NN)$ equations. If $x_0=1$ and $x_i=0$ for $i<0$, the $z_{im}$
with $(i,m)\in V(A_\NN,A_\NN)$ satisfy the $(A_\NN,A_\NN)$ equations. If in addition $x_{r+1}=1$ and $x_i=0$ for $i>r+1$, the $z_{im}$
with $(i,m)\in V(A_r,A_\NN)$ satisfy the $(A_r,A_\NN)$ equations.
\end{theorem}

\begin{proof}
For $A_\ZZ$ this is an immediate consequence of Jacobi's determinant identity. When the determinants are written as wedge products, this
identity states that
\begin{eqnarray}
&(v_1\wedge U\wedge v_2)\otimes (w_1\wedge U\wedge w_2)= \nonumber\\
&(v_1\wedge U\wedge w_2)\otimes (w_1\wedge U\wedge v_2)-(v_1\wedge U\wedge w_1)\otimes (w_2\wedge U\wedge v_2),
\end{eqnarray}
where $U=u_1\wedge\ldots\wedge u_{n-2}$ and $v_1,v_2,w_1,w_2,u_1,\ldots,u_{n-2}$ are vectors in $\CC^m$.

For the convenience of the reader we give a proof of this identity. One may assume that $U\not=0$. The equation is linear in the vectors
$v_i,w_i$. Thus it is sufficient to prove it for generic $w_1,w_2$, such that $v_1,v_2$ can be written as linear combinations of
$w_1,w_2,u_1,\ldots,u_{n-2}$. Linear combinations of the $u_i$ drop out, and the equation reduces to an identity in the two-dimensional
quotient space $\CC^m/\langle u_1,\ldots,u_{n-2}\rangle$, which is essentially the formula for the determinant of $2\times 2$ matrices. To
deduce the discrete Hirota equations one denotes the rows of $M^m_i$ by $v_1,u_1,\ldots,u_{n-2},v_2$ and puts $w_1=(1,0,\ldots,0)$,
$w_2=(0,\ldots,0,1)$.

If $x_0=1$ and $x_i=0$ for $i<0$, the matrices $M^m_0$ are upper triangular with determinant $z_{0m}=1$, such that the relevant components of
the $(A_\ZZ,A_\NN)$ equations reduce to the $(A_\NN,A_\NN)$ equations. If $x_{r+1}=1$ and $x_i=0$ for $i>r+1$, the matrices $M^m_{r+1}$ are
lower triangular with determinant $z_{r+1,m}=1$, as required for the $(A_r,A_\NN)$ equations. \qed
\end{proof}

\begin{theorem}\label{A2}
The map $R_1:S(A_r,A_\NN)\mapsto \CC^r$ is an isomorphism. For any $z\in S(A_r,A_\NN)$ one has $z_{im}=\det M^m_i(x)$, where $x_i=z_{i1}$ for
$i=1,\ldots,r$, $x_0=x_{r+1}=1$, and $x_i=0$ otherwise.
\end{theorem}

\begin{proof}
Let $z\in S^*(A_r,A_\NN)$. Since $R_1\circ M$ is the identity on $\CC^r$, one has $R_1MR_1z=R_1z$. By theorem~\ref{A1} and
equation~(\ref{X,Y}) this implies the set of polynomial equations $MR_1z=z$, which by definition are satisfied by all elements of
$S(A_r,A_\NN)$. Thus $R_1$ yields an isomorphism from $S(A_r,A_\NN)$ to a subset of $\CC^r$.

Moreover, $R_1(S^*(A_r,A_\NN))$ consists of those vectors  in $\CC^r$ for which none of the determinants $\det M^m_i(x)$ vanishes. This set
is dense in $\CC^r$, thus unconstrained by any polynomial equation. This implies that $S(A_r,A_\NN)$ is the solution set of the equations
$MR_1z=z$, which is isomorphic to $\CC^r$. \qed
\end{proof}

Using equation~(\ref{X,Y}) one obtains

\begin{corollary}
There is a unique injective map $\phi_{rn}:S^*(A_r,A_n)\rightarrow S(A_r,A_\NN)$ such that $\left(\phi_{rn}(z)\right)_{im}=z_{im}$ for
$(i,m)\in V(A_r,A_n)$, namely $\phi_{rn}=M\circ R_1$. Moreover, $\left(\phi_{rn}(z)\right)_{i,n+1}=1$ for $i\in V(A_r)$.
\end{corollary}

\begin{theorem}\label{A3}
For $z\in S(A_r,A_\NN)$ one has $z_{im}=\det M^i_m(y)$, where $y_k=z_{1k}$ for $k>0$, $y_0=1$ and $y_k=0$ for $k<0$. Moreover,
$\det M^{r+1}_m(y)=1$.
\end{theorem}

\begin{proof}
Let $z\in \CC^{V(A_r,A_\NN)}$ and $x_i=z_{i1}$. By theorem~\ref{A1}, the component $z_{im}$ is a polynomial of degree $m$ in $x_i$, with
leading term $x_i^m$, thus not identically zero. Thus one can apply equation~(\ref{X,Y}) and theorem~\ref{A1}, as in the proof of the
preceding theorem. \qed
\end{proof}

\begin{theorem}\label{A4}
Let $x_i=z_{i1}$ for $i=1,\ldots,r$ and $x_0=x_{r+1}=1$. Put $z_{10}=1$ and $z_{1,-n}=0$ for $n=1,\ldots,r$. Then for any $n\in\NN$ one has
\begin{equation*}
\sum_{k=0}^{r+1}(-)^kx_k z_{1,n-k}=0.
\end{equation*}
\end{theorem}

\begin{proof}
This is just the expansion of the matrix $M^n_1$ in terms of minors with respect to the elements of the last column. \qed
\end{proof}

\begin{theorem}\label{A5}
Let $g_1,\ldots,g_{r+1}$ be the roots of the polynomial
\begin{equation*}
p(\gamma) = \sum_{k=0}^{r+1}(-)^k x_{r+1-k}\gamma^k=0,
\end{equation*}
in some order, and put $g = diag(g_1, \ldots, g_{r+1})$. One has $g\in SL(r+1)$. Let $d(r+1)$ be the subgroup of $SL(r+1)$ consisting of the
diagonal matrices and let $\lambda_1,\ldots,\lambda_r$ be the fundamental highest weights of $SL(r+1)$ with respect to the maximal torus
$d(r+1)$. Let $\chi_{n\lambda_i}(g)$ be the value at $g$ of the character in the irreducible representation with highest weight $n\lambda_i$.
Then
\begin{equation*}
z_{1n}=\chi_{n\lambda_1}(g).
\end{equation*}
\end{theorem}

\begin{proof}
Consider an element of $S(A_r,A_\NN)$ for which $g$ is generic. The sequence $z_{1n}$ is a linear combination of the sequences $(g_i)^n$,
with $i = 1,\ldots, r+1$. Since $g_1\cdots g_{r+1}=1$, one has $g\in SL(r+1)$. Note that the Weyl group of $SL(r+1)$ with respect to
$d(r+1)$ is the permutation group of the diagonal entries, thus isomorphic to the permutation group $\mathcal{S}^{r+1}$ on the set
$\{1,\ldots,r+1\}$. The map from $(x_1,\ldots,x_r)$ to $g$ yields a bijection between $S(A_r,A_\NN)$ and the quotient of $d(r+1)$ by its
Weyl group. Recall the following cases of Weyl's character formula for diagonal matrices. For $n\in\ZZ$ let
\begin{equation}
N_{ni}(g) = \sum_{\omega\in\mathcal{S}^{r+1}} \mathrm{sgn}(\omega)
\prod_{j=1}^i g_{\omega(j)}^{r+n+1-j} \prod_{j=i+1}^r g_{\omega(j)}^{r+1-j}.
\end{equation}
The Weyl denominator is given by $D=N_{0i}$ independently of $i$. Then
\begin{equation}
\chi_{n\lambda_i}(g) = N_{ni}(g)/D(g).
\end{equation}
In our case the Weyl denominator formula takes the Vandermonde form
\begin{equation}
D(g) = \prod_{i<j} (g_i - g_j).
\end{equation}
For generic $g$ one has $D(g)\neq 0$ such that the Weyl character formula implies that $z_{1n}=N_{n1}(g)/D(g)$ satisfies the recursion
relation in Theorem 5. Moreover $z_{1n}=N_{n1}(g)/D(g)$ is true for $-n=0,\ldots,r$. Due to $x_{r+1}=1$ the recursion relation has a unique
solution with this property. Since we work on an irreducible variety, the restriction to $D\neq0$ is irrelevant. \qed
\end{proof}

\begin{theorem}[Kirillov and Reshetikin] \label{A6}
\begin{equation*}
z_{im}=\chi_{m\lambda_i}(g).
\end{equation*}
\end{theorem}

\begin{proof}
This is an immediate consequence of theorems~\ref{A3} and~\ref{A5}, together with the character formula $\chi_{m\lambda_i}(g)=\det M^i_m(y)$,
where $y_k=\chi_{k\lambda_1}(g)$. The latter was obtained by Schur in his PhD thesis~\cite{Schur}. Sometimes it is erroneously called Weyl's
second character formula. \qed
\end{proof}

As a preparation for the more involved proof for $X=D_r$ we rederive Schur's result by an explicit calculation using Weyl's character formula.

\begin{proof}
If one sums over all permutations $\omega$ with fixed $\omega(1),\ldots,\omega(i)$ and uses the Weyl denominator formula for the Weyl group
of $SL(r+1-i)$ one obtains
\begin{equation}
\chi_{n\lambda_i} = \sum_{\sigma \in I(i,r+1)} D_{\sigma}(g)^{-1} \prod_{j=1}^i g_{\sigma(j)}^{n+r+1-j},
\end{equation}
where $I(i,r+1)$ is the set of all injective maps from $\{1,\ldots,i\}$ to $\{1, \ldots, r+1\}$ and $D_{\sigma}=D_{\sigma}^1 D_{\sigma}^2$,
with
\begin{eqnarray}
D_{\sigma}^1 &=& \prod_{1\leq j<k\leq r+1}\left(g_{\sigma(j)} -g_{\sigma(k)}\right),\\
D_{\sigma}^2 &=& \prod_{j=1}^i \prod_{k \in \sigma^c}\left(g_{\sigma(j)}-g_k\right),
\end{eqnarray}
where $\sigma^c$ is the complement of the image of $\sigma$ in $\{1,\ldots,r+1\}$.
In particular we have $z_{1n} = \sum_{j=1}^{r+1}\phi_{jn}$ with
\begin{equation}
\phi_{jn} = g_j^{n+r}/\prod_{k\neq j}(g_j-g_k).
\end{equation}

Thus
\begin{equation}
z_{in} = \sum_{\mu\in I(i,r+1)}\det M_{\mu}^n,
\end{equation}
where $M_{\mu}^n$ is the $i\times i$ matrix with elements
\begin{equation}
(M_{\mu}^n)_{kl} = \phi_{\mu k, l-k+n}.
\end{equation}
Non-injective maps do not contribute since they lead to matrices with linearly dependent rows. The numerators of the $M_{\mu}^n$ matrix elements form a Vandermonde type matrix with determinant
\begin{equation}
\prod_{j=1}^{i}g_{\mu j}^{n+r-j+1}\prod_{1\leq k <l\leq i}(g_{\mu l}-g_{\mu k}).
\end{equation}
This yields
\begin{equation}
z_{in} = \sum_{\mu\in I(i, r+1)}\prod_{1\leq k<l\leq1}(g_{\mu l}-g_{\mu k})^{-1}\prod_{j=1}^i\left( g_{\mu j}^{n+r-j+1}
\prod_{s\in C(\mu)}(g_{\mu j} - g_s)^{-1} \right),
\end{equation}
where $C(\mu)$ is the complement of the image of $\mu$ in $\{1, \ldots, r+1\}$. For given $\mu$ let $N(\mu)$ be the set of all injective maps
from $\{i+1,\ldots,r\}$ to $C(\mu)$. We have the Vandermonde determinant formula
\begin{equation}
\sum_{\nu\in N(\mu)}\prod_{j=i+1}^r g_{\nu j}^{r-j+1}\prod_{i\leq k<l\leq r}(g_{\nu l}-g_{\nu k})^{-1} = 1.
\end{equation}
Inverting the left hand side in the previous formula and noting that the pairs $\mu$ and $\nu$ with $\nu\in N(\mu)$ form the permutation
group $W(r+1)$ of $\{1, \ldots, r\}$, we obtain
\begin{equation}
z_{in} = \sum_{\sigma\in W(r+1)}\prod_{1\leq k<l\leq r}(g_{\sigma k} - g_{\sigma l})^{-1} \prod_{j=1}^i g_{\sigma j}^{n+r-j+1}
\prod_{k=i+1}^r g_{\sigma j}^{r-k+1}.
\end{equation}
Note that $z_{in}$ has only single poles at the locus where two $g_i$ coincide. We shall refer to this subset of $U(r+1)$ as the singularity
locus. For all $\sigma\in W(r+1)$ we have
\begin{equation}
\prod_{1\leq k<l\leq r}(g_{\sigma k} - g_{\sigma l}) = \mathrm{sgn}(\sigma)D(A_r),
\end{equation}
where $D(A_r)$ is the Weyl denominator. Thus we have
\begin{equation}
D(A_r)z_{in} = \sum_{\sigma\in W(r+1)}\mathrm{sgn}(\sigma) \prod_{j=1}^i g_{\sigma j}^{n+r+1-j} \prod_{j = i+1}^r g_{\sigma j}^{r+1-j},
\end{equation}
or according to the Weyl character formula
\begin{equation}
z_{in} = \chi_{n \lambda_i}(g).
\end{equation}
\qed
\end{proof}

\begin{remark}
The previous formula can be written somewhat more concisely as follows. We regard characters as polynomials in $g_1, \ldots, g_{r+1}$ and
introduce the natural action of $\mathcal{S}^{r+1}$ on the ring of polynomials by $\sigma \left(g_i\right) = g_{\sigma (i)}$. With
\begin{equation}
\mathcal{A}_r = \sum_{\sigma\in {\mathcal{S}^{r+1}}} \left(\mathrm{sgn}(\sigma)\right) \sigma,
\end{equation}
we have
\begin{equation}
D(A_r)z_{in} = \mathcal{A}_r \prod_{j=1}^i g_j^{n+r+1-j} \prod_{j=i+1}^r g_j^{r+1-j}.
\end{equation}
\end{remark}

\vspace{5mm}
So far we have reproduced known results. We now consider $S(A_r,A_\ZZ)$.

\begin{theorem}\label{A7}
$S(A_r,A_\ZZ)$ is birationally isomorphic to the product of two maximal tori of $SL(r+1)$ modulo the simultaneous action of the Weyl group on
the two factors. Explicitly, let $g=\mathrm{diag}(g_1,\ldots,g_{r+1})$ and $h=\mathrm{diag}(h_1,\ldots, h_{r+1})$ be elements in the torus
and put
\begin{equation*}
N_{ni}(g,h)=\sum_{w\in \mathcal S^{r+1}}\mathrm{sgn}(w)\prod_{j=1}^i g_{w(j)}^{r+n+1-j}h_{w(j)}^{r+1-j}
\prod_{j=i+1}^r g_{w(j)}^{r+1-j}.
\end{equation*}
For $D(g)\neq 0$ the corresponding element of $S(A_r,A_\ZZ)$ is given by
\begin{equation*}
z_{in}=N_{ni}(g,h)/D(g). \label{WCF}
\end{equation*}
\end{theorem}

\begin{proof}
For generic $g\in SU(r+1)$, the group elements $g^n$, $n$ a positive integer, are dense in $SU(r+1)$. Thus any algebraic relation between the
diagonal matrix elements of $g$ and $g^n$ which is true for arbitrary positive integers $n$ stays true when $g^n$ is replaced by an arbitrary
diagonal matrix $h\in SU(r+1)$. By complexification the result stays true for $SL(r+1)$. The algebraic relations between the $z_{im}$ of
$S(A_r,A_\NN)$ remain true when the $z_{im}$ are replaced by $z_{i,m+n}$ for arbitrary positive integers $n$. Thus the formula for $z_{in}$
given above yields elements of $S(A_r,A_\ZZ)$. We conclude by comparing dimensions. \qed
\end{proof}

The case $h=1$ yields
\begin{corollary}
There is a natural embedding $S(A_r,A_\NN)\subset S(A_r,A_\ZZ)$.
\end{corollary}

\begin{theorem}\label{A7}
$S(A_r,A_\ZZ)$ is fibered over a maximal torus of $SL(r+1)$ modulo the action of the Weyl group.
\end{theorem}

\begin{proof}
By theorems~\ref{A1} and~\ref{A3}, any $z\in S(A_\NN,A_\ZZ)$ given by $z_{im}=\det M^i_m(y)$ satisfies the $(A_\NN,A_\ZZ)$ equations. In
addition, $z_{r+1,m}=1$ for $z\in S(A_r,A_\ZZ)$, which implies $z_{r+2,m}=\det M^{r+2}_m(y)=0$. This means that the vectors
$v_m$, $v_{m+1}$,$\ldots$, $v_{m+r+1}$ with  $v_m=(y_m,y_{m+1},\ldots,y_{m+r+1})$ are linearly dependent for any $m$. On the other hand,
$\det M^{r+1}_m(y)=1$ implies that there is no linear dependency between $v_{m+1}\ldots,v_{m+r+1}$. This means that for each
$z\in S(A_\NN,A_\ZZ)$ there is a unique relation $\sum_{k=0}^{r+1} (-)^k a_k v_{m+k}=0$ with $a_0=1$. When $z$ is given by $g,h$ with
$D(g)\neq 0$, the $a_k$ are given by theorem \ref{A5} in terms of $g$, independently of $m,h$. Algebraic relations in $S(A_r,A_\ZZ)$ which
hold for generic elements hold for all of $S(A_r,A_\ZZ)$. Thus the $a_k$ are always independent of $m$ and $a_{r+1}=1$. As in theorem
\ref{A5}, this yields a unique element of the maximal torus of $SL(r+1)$ modulo the action of the Weyl group. \qed
\end{proof}

Now we determine the elements of $S(A_r,A_n)$.

\begin{theorem}\label{A8}
Let $z$ be the image of an element of $S(A_r,A_n)$ in $(A_r,A_\NN)$. Then $z_{1m}=0$ for $m=n+2,n+3,\ldots,n+1+r$.
\end{theorem}

\begin{proof}
By assumption we have $z_{i,n+1}=1$ for all $i$. The $(A_r,A_\NN)$ equations immediately yield $z_{i,n+2}=0$ for all $i$. Now we use
induction on $m$. Assume that $z_{1k}=0$ for $k=n+2,\ldots,n+m$. By theorem \ref{A3} we have $z_{im}=\det M^i_m(y)$, where $y_k=z_{1k}$. The
minors of $\det M^i_m(y)$ with respect to the first row have vanishing determinant, except for the last one, which yields 1. Thus
\begin{equation}
z_{m,n+2}=(-)^{m+1}z_{1,n+m+1}.
\end{equation} \qed
\end{proof}

\begin{theorem}\label{A9}
\begin{equation*}
z_{1,m+n+2+r}=(-1)^r z_{1m}.
\end{equation*}
\end{theorem}

\begin{proof}
For $m=-r,\ldots,-1$ this is true by the previous theorem. For $m=0$ it follows from that theorem by the recursion relation in Theorem 5. For
larger $m$ it follows from the recursion relation by induction. \qed
\end{proof}

\begin{theorem}\label{A10}
Let $d(r+1)$ be the diagonal subgroup of $SU(r+1)$. The elements of $S(A_r,A_n)$ are in one-to-one correspondence with the set of elements
$g\in d(r+1)$ satisfying $g^{r+n+2}=(-1)^r$, where all eigenvalues of $g$ are different, modulo permutations of the eigenvalues.
\end{theorem}

\begin{proof}
By l'Hopital's rule we can write for given $g$
\begin{equation}
\chi_{m\lambda_1} = \sum_{i=1}^{r+1}p_i(m)(g_i)^m,
\end{equation}
where the degree of the polynomial $p_i$ is bounded by the number of $g_j$ with $g_j=g_i$, $i\neq j$. To have well-defined $p_i$ we put
$p_j=0$ for $g_j=g_i$ with $j>i$. The natural continuation formula for $\chi_{m\lambda_1}(g)$ to negative $m$ agrees with the continuation
of $z_{1m}$ by the recursion relation in Theorem 5. For large positive $m$ the value of $\chi_{m\lambda_1}$ is dominated by the contribution
of the $g_i$ with largest absolute value and highest power of $m$ among the corresponding $p_i$. Analogously, the behaviour for large
negative $m$ is dominated by the contribution of the $g_i$ with smallest absolute values. From the periodicity of $z_{1m}$ it follows that
$\chi_{m\lambda_1}(g)$ remains bounded, such that $g\in d(r+1)$ and one can use constant $p_i$. Periodicity in $m$ yields $g^{n+r+2}=1$. The
$r+1$ vectors $V_m=\left(z_{1m},z_{1,m+1},\ldots,z_{1,m+r}\right)$ for $m=n+1,n,\ldots,n-r$ are linearly independent, since the matrix with
row vectors $V_{n+1},V_n,\ldots,V_{n-r}$ is upper triangular with non-vanishing diagonal. Thus the $r+1$ vectors $1, g_i, \ldots,(g_i)^r$ for
$i=0,\ldots,r+1$, must be linearly independent too. This implies that the $g_i$ are pairwise different. \qed
\end{proof}

\section{The D Case}
\label{sec:2}

Now let us consider $S(D_r,A_n)$. We choose a real form of $D_r$ such that a maximal torus of the corresponding orthogonal group is
parametrised as $g=\mathrm{diag}(g_1,\ldots,g_{2r})$ with $g_{i+r}=g_i^{-1}$ for $i=1,\ldots,r$. We label the vertices of the $D_r$ Dynkin
diagram such that the nodes $r-1$ and $r$ correspond to the half-spinor representations. The remaining nodes are numbered consecutively from
$1$ to $r-2$, starting at the outer vertex which corresponds to the vector representation.

Again we first consider $S(D_r,A_\NN)$. As for $S(A_r,A_\NN)$ the generic solutions of the discrete Hirota equations form an affine variety
parametrised by $z_{i1}$. The formulas are slightly more complicated and were given in \cite{KNH}.

\begin{theorem}[Kuniba, Nakamura and Hirota]\label{D1}
Given complex numbers $x_0$, $x_1$,$\ldots$, $x_r$ with $x_0=1$, and $j\in\ZZ$, let $X_j$ and $Y_j$ be defined by
\begin{eqnarray*}
X_{2j+1}=x_j=-X_{4r-3-2j} &{\mathit for}& 0\leq j \leq r-2\\
X_{2j}=-x_j &{\mathit for}& j=r-1,r\\
X_j=0 &{\qquad \mathit otherwise}&
\end{eqnarray*}
\begin{equation*}
Y_2=-Y_6=1,\qquad Y_3=x_r,\qquad Y_5=x_{r-1}
\end{equation*}
\begin{equation*}
Y_j=0 \qquad {\mathit otherwise}.
\end{equation*}
We will consider square matrices $T^a_k$ of size $2k-1$ for $a=1,\ldots,r-2$, and of size $2k$ for $a=r-1,r$, where $k=1,2,\ldots$. Their
matrix elements have the property
\begin{equation*}
(T^a_k)_{ij}=(T^a_k)_{i+2,j+2}
\end{equation*}
if both sides of the equation are defined. Moreover,
\begin{eqnarray*}
(T^a_k)_{1j}&=X_{2a+j}\\
(T^a_k)_{2j}&=Y_j
\end{eqnarray*}
for $a=1,\ldots,r-2$, whereas
\begin{eqnarray*}
(T^{r-1}_k)_{1j}&=&Y_{3+j}\\
(T^{r-1}_k)_{2j}&=&X_{2r-3+j}\\
(T^r_k)_{1j}&=&X_{2r-2+j}\\
(T^r_k)_{2j}&=&Y_{2+j}.
\end{eqnarray*}
Note that $T^a_k$ is antisymmetric for $a=r-1,r$, such that it has a Pfaffian. Put $x_i=z_{i1}$ for $i=1,\ldots,r$. Then in $S(D_r,A_n)$ one
has
\begin{equation*}
z_{ak}=\det \left(T^a_k\right),
\end{equation*}
for $a=1,\ldots,r-2$ and
\begin{equation*}
z_{ak}=\mathrm{Pf}\left(T^a_k\right),
\end{equation*}
for $a=r-1,r$.
\end{theorem}

\begin{proof}
The proof is again an immediate consequence of Jacobi's identity for suitably chosen matrices, together with the standard relation between
determinants and Pfaffians. \qed
\end{proof}

\begin{remark}
Consequently  $S(D_r,A_\NN)$ has properties analogous to $S(A_r,A_\NN)$. More specifically, $S(D_r,A_\NN)$ is isomorphic to $\CC^r$, with
elements parametrised by arbitrary vectors $(z_{11},\ldots,z_{r1})$, and the map $\phi_{rn}:S(D_r,A_n)\rightarrow S(D_r,A_\NN)$, induced by
the restriction to
$(z_{11},\ldots,z_{r1})$, is an embedding.
\end{remark}

We now derive a recursion relation for the $z_{1m}$.

\begin{theorem}\label{D2}
Let $x_i=z_{i1}$ for $i=1,\ldots,r$ and $x_0=x_{r+1}=1$. Put $z_{10}=1$ and
$z_{1,-n}=0$ for $n=1,\ldots,2r-1$. Then for any $n\in\NN$ one has
\begin{equation*}
\sum_{k=0}^{2r}(-)^k \chi_k z_{1,n-k}=0,
\end{equation*}
where $\chi_{2r-k}=\chi_k$ for $k=0,\ldots,r$ and
$\chi_0= 1$, $\chi_1=x_1$, $\chi_i=x_i-x_{i-2}$ for $i=2,\ldots,r-2$,
$\chi_{r-1}=x_{r-1}x_r-x_{r-3}$ and $\chi_r=x_{r-1}^2+x_r^2-2x_{r-2}$.
\end{theorem}

\begin{proof}
Expanding the minors of $\det(T^1_k)$ with respect to the first row one finds
\begin{eqnarray}
&& \sum_{i=0}^{r-2} (-)^i x_i \left(\det \left(T^1_{k-i}\right)-\det\left(T^1_{k-2r+2+i}\right)\right) \nonumber\\
&+& x_{r-1}\det\left(\tilde T_{2k-2r+3}\right)-x_r\det\left(\tilde T_{2k-2r+1}\right)=0,
\end{eqnarray}
and
\begin{equation}
\det\left(\tilde T_{2k+3}\right)-x_r \det\left(T^1_{k+1}\right)-x_{r-1}\det\left(T^1_k\right)-\det\left(\tilde T_{2k-1}\right)=0.
\end{equation}
Here $\tilde T_{2k-2r+3}$ is the matrix obtained by suppressing rows $1,\ldots,2r-5$ and $2r-3$ and columns $1,\ldots,2r-4$ in $T^1_k$. For
small values of $k$ the recursion relations are to be interpreted so that $\det \left(T^1_{-k}\right)=0$ for $k=1,\ldots,2r-3$,
$\det \left(T^1_{-2r+2}\right)=1$ and $\det \left(\tilde T_{-2k+1}\right)=0$ for $k=1,\ldots,r$. Eliminating the $\tilde T$-determinants by
subtracting the recursion relations for $\det\left(T^1_k\right)$ and $\det\left(T^1_{k-2}\right)$ yields
\begin{equation}
\sum_{i=0}^{r-1} (-)^i \chi_i \left(\det\left(T^1_{k-i}\right)+
\det\left(T^1_{k-2r+i}\right)\right)+(-)^r\chi_r\det\left(T^1_{k-r}\right) =0.
\end{equation} \qed
\end{proof}

\begin{remark}
The solutions of the recursion relations have the form
\begin{equation}
z_{1k}=\sum_{i=1}^{2r} \alpha_i g_i^k,
\end{equation}
where the $g_i$ are the solutions of the polynomial equation in $\gamma$
\begin{equation}
\sum_{i=0}^{r-1} (-)^i \chi_i (\gamma^i+\gamma^{2r-i})+(-)^r\chi_r \gamma^r =0\ ,
\end{equation}
and one may put $g_{i+r}=g_i^{-1}$ since the polynomial coefficients are invariant under reversal of their order. Expressing $\chi_i$ in
terms of $g_1,\ldots,g_r$ one notes that it can be interpreted as the character of the $i$-fold antisymmetric product of the fundamental
representation of $D_r$, evaluated at the group element $g=\mathrm{diag}(g_1,\ldots,g_{2r})$. This implies that $x_{r-1}$ and $x_r$ can be
interpreted as characters of the half-spin representations and that all $x_i$ can be understood as Yangian characters.
\end{remark}

\begin{theorem}\label{D3}
Let $\{g_1,\ldots,g_{2r}\}$ with $g_{i+r}=g_i^{-1}$ for $i=1,\ldots,r$ be the solution set of the polynomial equation in $\gamma$
\begin{equation*}
\sum_{i=0}^{2r} (-)^i \chi_i \gamma^i =0.
\end{equation*}
Then
\begin{equation*}
z_{1k}=\chi_{k\lambda_1}(g).
\end{equation*}
\end{theorem}

\begin{proof}
The proof is analogous to the $A_r$ case. Let
\begin{equation}
N_k(g) =\sum_{\epsilon\in\{-1,1\}^{r-1}}\sum_{\sigma\in {\mathcal S}^r} \mathrm{sgn}(\sigma)
g_{\sigma(1)}^{\epsilon_1(r-1+k)}g_{\sigma(2)}^{\epsilon_2(r-2)}g_{\sigma(3)}^{\epsilon_3(r-3)}\cdots
g_{\sigma(r-1)}^{\epsilon_{r-1}}.
\end{equation}
Then the Weyl character formula can be written as $\chi_{k\lambda_1}(g)=N_k(g)/N_0(g)$. Clearly the $N_k$ satisfy the recursion formula, and
one also has $N_{-k}(g)=0$ for $k=1,\ldots,2r-3$, and $N_{-2r+2}(g)=1$. Thus $N_k(g)/N_0(g)=z_{1k}$ for $k=-2r+2,\ldots,1$, which implies
that the same relation holds for all $k$ due to the recursion relation. \qed
\end{proof}

For completeness we use the preceding result to give a direct derivation of the Kirillov-Reshetikhin formula for $D_r$.
\begin{theorem}\label{D4i}
For $i=1,\ldots,r-2$
\begin{equation*}
z_{in} = \sum{K(i,n)}\chi_{\sum_{j=1}^i k_j\lambda_j}(g),
\end{equation*}
where for odd $i$ the set $K(i,n)$ consists of the tuples $k_1,\ldots,k_i$ such that $k_j=0$ for even $j$ and
\begin{equation*}
\sum_{j=1}^i k_j=n,
\end{equation*}
whereas for even $i$ the set $K(i,n)$ consists of the tuples $k_1,\ldots,k_i$ such that $k_j=0$ for odd $j$ and
\begin{equation*}
\sum_{j=2}^i k_j \leq n.
\end{equation*}
\end{theorem}

\begin{proof}
We put
\begin{equation}
Z_{in}=\det M^i_n,
\end{equation}
where $i=1,\ldots,r$ and $M^i_n$ is the $i\times i$ matrix with elements
\begin{equation}
\left(M^i_n\right)_{jk}=z_{1,n-j+k}.
\end{equation}
For $i = 1, \ldots, r-2$ the Dynkin diagrams of $A_r$ and $D_r$ are identical. Therefore we must have $Z_{in}=z_{in}$ for $i=1,\ldots,r-2$.
On the other hand, the relation $z_{r-2,n}^2 = z_{r-2,n-1}z_{r-2,n+1} + z_{r-3,n}z_{r-1,n}z_{r,n}$ yields
\begin{equation}
Z_{r-1,n} = z_{r-1,n}z_{rn},
\end{equation}
and a short calculation shows that the relations for $i = r-1,r$ yield
\begin{equation}
Z_{rn} = z_{rn}^2 + z_{r-1,n}^2 - z_{r-2,n}.
\end{equation}
We have seen that $z_{1n}=\chi_{n\lambda_1}(g)$, where $g$ is a diagonal $SO(r,r)$ matrix with entries $g_j, g_j^{-1}$, $j=1,\ldots,r$. We
again refer to the set where the two $g_i$ coincide as the singularity locus. Since the $z_{in}$ are invariant under permutation of the
$g_i$ and under the replacement of any $g_i$ by its inverse, the same is true for the $Z_{in}$. Permutations and replacements of an even
number of the $g_i$ by their inverses generates the Weyl group of $SO(r,r)$. On the other hand the single flip $\xi$ defined by
$\xi g_r = g_r^{-1}$, $\xi g_i = g_i$ for $i<r$ yields an outer automorphism which fixes the fundamental weights $\lambda_i$ for
$i=1,\ldots,r-2$, but interchanges $\lambda_r$ and $\lambda_{r-1}$.

By construction, the $Z_{in}$ are $\ZZ$-linear combinations of $SO(r,r)$ characters and in particular $\ZZ$-linear combinations of the
$SO(r,r)$ weight $\lambda$. One writes
\begin{equation}
g^\lambda=\prod_{i=1}^r g_i^{l_i},
\end{equation}
or equivalently $\lambda = (l_1,\ldots, l_r)$, where the $l_i$ are either all integral or all half integral. For later use we need a notation
which suppresses the dependence on $g_1,\ldots,g_k$ with $k\in\{0,\ldots,r\}$. We write
\begin{equation}
g^{\lambda}[k] = \prod_{i=k+1}^r g_i^{n_i}.
\end{equation}
Dominant weights are those for which $n_1 \geq n_2 \geq \ldots \geq n_{r-1} \geq |n_r|$. We will use the weights
\begin{equation}
\Lambda_i = (\underbrace{1,\ldots,1}_{i},0,\ldots,0),
\end{equation}
for $i=1,\ldots,r$. In terms of the fundamental weight $\lambda_i$ one has $\Lambda_i=\lambda_i$ for $i\leq r-2$,
$\Lambda_{r-1} = \lambda_{r-1} + \lambda_r$ and $\Lambda_r = 2\lambda_r$. Recall that $\ZZ[x]$ is the ring of polynomials in $x$ with integer
coefficients and $\ZZ(x)$ is the corresponding ring of power series. By abuse of notation we denote by $\ZZ[x_I]$ the ring of polynomials in
$x_i$ with $i\in I$ and  analogously for $\ZZ(x_I)$. Instead of $\ZZ[x](y)/\langle xy-1 \rangle$ we use the abbreviated notation
$\mathcal{R}[x](x^{-1})$.

Let $P\in \mathrm{End}\ZZ[g_1,\ldots,g_r](g_1^{-1},\ldots,g_r^{-1})$ be the projection to the dominant weight and $P_{\geq}$, $P_0$,
$P_{\leq}$ the subprojections to dominant weights with $n_r\geq0$, $n_r=0$, and $n_r\leq0$ respectively. In particular,
$P_{\leq} = \xi P_{\geq} \xi$. We use $P[k]$ to denote the projection to the span of $g^{\lambda}[k]$, with $\lambda$ any dominant weight,
and analogously for $P_{\geq}[k]$.

When $\chi_{\lambda}$ is the character of an irreducible representation with highest weight $\lambda$ and $\Delta_r$ is the Weyl denominator,
we have
\begin{equation}
g^{\lambda} = P g^{-\rho}\Delta_r \chi.
\end{equation}
Here $\rho=\sum_{i=1}^r \lambda_i$, such that
\begin{equation}
g^\rho=\prod_{i=1}^r g_i^{r-i}.
\end{equation}
Since $\xi Z_{in}=Z_{in}$, one has
\begin{equation}
Pg^{-\rho}\Delta_r Z_{in} = (1 + \xi - P_0)P_{\geq}g^{-\rho}\Delta_r Z_{in},
\end{equation}
so it is sufficient to calculate $P_{\geq}g^{-\rho}\Delta_r Z_{in}$ to obtain the decomposition of $Z_{in}$ into characters of irreducible
representations.

We write $\Delta_r=\Delta_r^+\Delta_r^-$, where
\begin{equation}
\Delta_r^+ = \prod_{1\leq i<j\leq r}(g_i - g_j),
\end{equation}
and
\begin{equation}
\Delta_r^- = \prod_{1\leq i<j\leq r}(1-g_i^{-1}g_j^{-1}).
\end{equation}
Using $(1-x)^{-1} = 1+x+x^2+\ldots$ we can regard the inverse of $\Delta_r^-$ as an element of $\ZZ(g_1^{-1},\ldots,g_r^{-1})$.

By the Weyl character formula for $D_r$ and the Weyl denominator formula for $D_{r-1}$ we have
\begin{equation}
z_{1n}=\sum_{j=1}^r \phi_j^n + \phi_-^n,
\end{equation}
where
\begin{equation}
\phi_{jn}=g_j^{n+r-1}/\prod_{k\not= j} (g_j-g_k)(1-g_j^{-1}g_k^{-1}),
\end{equation}
and
\begin{equation}
\phi_-^n = \sum_{j=1}^r g_j^{-(n+r-1)}/\prod_{k\neq j}\left((g_j - g_k)(1-g_j^{-1}g_k^{-1})\right).
\end{equation}
In $\phi_-^n$ the poles at the singularity locus are removable, so that
\begin{equation}
\phi_-^n \in \prod_{j=1}^r g_j^{-1}\ZZ(g_1^{-1},\ldots,g_r^{-1}).
\end{equation}
One has
\begin{equation}
Z_{in} = \sum_{\mu\in\nu(i,r)}\det M^{\mu}_n,
\end{equation}
where $\nu(i,r)$ is the set of maps from $\{1, \ldots,i\}$ to $\{1,\ldots,r,-\}$ and the $i\times i$ matrices $M^{\mu}_n$ have matrix
elements
\begin{equation}
\left(M^{\mu}_n\right)_{jk} = \phi_{\mu j}^{n-j+k}.
\end{equation}
For a given map $\mu$ let $n(\mu)$ be the subset of $\{1,\ldots,i\}$ which is mapped to $-$, and let $p(\mu)$ be its complement. If $\mu$ is
not injective on $p(\mu)$ then $\det M^{\mu}_n=0$. Otherwise $\det M^{\mu}_n$ has only single poles at the singularity locus. The development
into minors with respect to $p(\mu)$ yields
\begin{equation}
\Delta_r^+\det M^{\mu}_n \in \Delta_{\mu}^c \prod_{i=1}^r g_i^{-|n(\mu)|}\ZZ[g_{\mu p(\mu)}]\left(g_1^{-1},\ldots,g_r^{-1}\right),
\end{equation}
where
\begin{equation}
\Delta_{\mu}^c = {\prod_{i,j\in c(\mu)}}_{i<j}(g_i - g_j),
\end{equation}
and $c(\mu)$ is the complement of $\mu p(\mu)$ in $\{1,\ldots,r\}$. Using the Weyl denominator formula for $\Delta_{\mu}^c$, we see that
every monomial in $\Delta_{\mu}^c\prod_{i=1}^r g_i^{-|n(\mu)|}$ has at least $|n(\mu)|$ strictly negative exponents for $g_i$ with
$i\in c(\mu)$. Thus
\begin{equation}
P_{\geq}g^{-\rho}\Delta_r\det M^{\sigma}_n=0,
\end{equation}
unless $n(\sigma)=\phi$. This implies
\begin{equation}
P_{\geq}g^{-\rho}\Delta_r Z_{in} = P_{\geq}g^{-\rho}\Delta_r\sum_{\sigma\in S(i,r)}\det M^{\sigma}_n,
\end{equation}
where $S(i,r)$ is the set of injective maps from $\{1,\ldots,i\}$ to $\{1,\ldots,r\}$. For $\sigma\in S(i,r)$ the calculation of
$\det M^{\sigma}_n$ proceeds as for $S(A_r,A_\NN)$ and yields
\begin{equation}
P_{\geq}g^{-\rho}\Delta_r Z_{in} = P_{\geq}g^{-\rho}\mathcal{A}(0,r)(N_i^0)^{-1}N_{r-i}^i g^{n\Lambda_i+\rho},
\end{equation}
where
\begin{equation}
N_{i-t}^t = \prod_{t<j<k\leq i}\left(1-g_j^{-1}g_k^{-1}\right),
\end{equation}
and
\begin{equation}
\mathcal{A}(t,s) = \sum_{\sigma\in\Pi(t,s)}\sigma \left(\mathrm{sgn}(\sigma)\right),
\end{equation}
and $\Pi(t,s)$ is the permutation group of $I(t,s)=\{t+1,\ldots,s\}$. Since all terms of $N_{r-i}^i$ except the identity yield terms which
project to zero, the preceeding formula can be simplified to
\begin{equation}
P_{\geq}g^{-\rho}\Delta_r Z_{in} = P_{\geq}g^{-\rho}\mathcal{A}(0,r)(N_i^0)^{-1}g^{n\Lambda_i+\rho}.
\end{equation}
Now
\begin{equation}
N_s^0 = N_{s-1}^1 \sum_{M\subset I(1,s)}(-)^{|M|}g_1^{-|M|}\prod_{a\in M}g_a^{-1},
\end{equation}
yields
\begin{equation}
(N_s^0)^{-1} = (N_{s-1}^1)^{-1} - (N_s^0)^{-1} \sum_{\substack{M\subset I(1,s)\\M\neq 0}}(-)^{|M|}g_1^{-|M|}\prod_{a\in M}g_a^{-1}.
\end{equation}
Moreover $\mathcal{A}(0,r)$ commutes with $N_s^0$ and factorises through $\mathcal{A}(1,r)$. When we put
\begin{equation}
G(t,s) = \prod_{t<a\leq s}g_a^{-a},
\end{equation}
we have
\begin{equation}
\mathcal{A}(1,s)G(1,s)\prod_{a\in M}g_a^{-1} = 0,
\end{equation}
unless $M$ is of the form $I(t,s)$ with $1\leq t\leq s$, since for $a\in M$, $a+1 \in M$, $a+1\leq s$ one gets equal exponents of $g_a$ and
$g_{a+1}$, thus zero when one averages over a group which contains the transposition of $a,\ a+1$. For $M=I(t,s)$ one obtains
\begin{equation}
\mathcal{A}(0,r)g_1^{t-s} \prod_{a\in I(t,s)}g_a^{-1} = 0 \quad\textrm{unless}\quad t=s/2,
\end{equation}
since for $t<s/2$ the exponents of $g_1$ and $g_{s-t}$, for $t>s/2$ those of $g_1$ and $g_{s-t+1}$, coincide. Thus the sum over $M$ yields no
contribution for odd $i$, whereas for even $i$ only the term $M=I(1,i/2)$ contributes. For odd $i$ this yields
\begin{eqnarray}
&& Pg^{-\rho}\mathcal{A}(0,r)(N_i^0)^{-1}g^{n\Lambda_i+\rho} \nonumber\\
&=& g_1^nP[1]g^{-\rho}[1]\mathcal{A}(1,r)(N_{i-1}^1)^{-1}g^{n\Lambda_i+\rho}[1],
\end{eqnarray}
and for even i
\begin{eqnarray}
&& Pg^{-\rho}\mathcal{A}(0,r)(N_i^0)^{-1}g^{n\Lambda_i+\rho} \nonumber\\
&=& g_1^nP[1]g^{-\rho}[1]\mathcal{A}(1,r)(N_{i-1}^1)^{-1}g^{n\Lambda_i+\rho}[1]\\
&+& Pg^{-\rho}\mathcal{A}(0,r)(N_i^0)^{-1}g^{(n-1)\Lambda_i+\rho}.\nonumber
\end{eqnarray}
By induction on $n$ and $k$ one obtains
\begin{eqnarray}
&& P_{\geq}[k]g^{-\rho}[k]\mathcal{A}(k,r)(N_{i-k}^k)^{-1}\prod_{j=k+1}^i g_j^{n+r-j}\prod_{j=i+1}^r g_j^{r-j} \nonumber\\
&=& \sum_{\lambda\in L(i,n,k)}g^{\lambda},
\end{eqnarray}
where $L(i,n,k)$ is the set of weakly decreasing integral sequences such that $l_{k+1}\leq r$, $l_m=0$ for $m>i$ and $l_{i-2j-1}=l_{i-2j}$
for $j=0,\ldots,[i/2]-1$. For $k=0$ this yields the desired decomposition of $Z_{in}$ in terms of characters of $SO(r,r)$. Equivalently one
can write for $i=1,\ldots,r-2$
\begin{equation}
z_{in} = \sum{K(i,n)}\chi_{\sum_{j=1}^i k_j\lambda_j}(g),
\end{equation}
where for odd $i$ the set $K(i,n)$ consists of the tuples $k_1,\ldots,k_i$ such that $k_j=0$ for even $j$ and
\begin{equation}
\sum_{j=1}^i k_j=n,
\end{equation}
whereas for even $i$ the set $K(i,n)$ consists of the tuples $k_1,\ldots,k_i$ such that $k_j=0$ for odd $j$ and
\begin{equation}
\sum_{j=2}^i k_j \leq n.
\end{equation}
 \qed
\end{proof}

For $i=r-1,r$ they stated the following result
\begin{theorem}\label{D4ii}
For $i=r-1,r$
\begin{equation*}
z_{in} = \chi_{n\lambda_i}(g).
\end{equation*}
\end{theorem}

\begin{proof}
We verify this result by proving that
\begin{equation}
Z_{r-1,n}=\chi_{n\lambda_{r-1}}(g)\chi_{n\lambda_r}(g),
\end{equation}
and
\begin{equation}
Z_{r,n} = \left(\chi_{n\lambda_{r-1}}(g)\right)^2 + \left(\chi_{n\lambda_r}(g)\right)^2 - z_{r-2,n}.
\end{equation}
The latter equations show that the pair $z_{r-1,n}$, $z_{r,n}$ agrees with $\chi_{n\lambda_{r-1}}(g)$, $\chi_{n\lambda_r}(g)$ in some order.
Once the order is fixed for $n=1$, the Kirillov-Reshetikhin result follows from $z_{r-1,n}z_{r,n-1}=z_{rn}^2 -z_{r-2,n}$ by induction on $n$
and comparison of the highest powers of $g_n$.

To prove the statements note that
\begin{equation}
Pg^{-\rho}\Delta_r\chi_{\lambda}\chi_{n\lambda_r} = Pg^{-\rho}\chi_{\lambda}\mathcal{A}(0,r) \sum_{\epsilon\in E^+}\prod_{i=1}^r
g_i^{\epsilon_i(n/2+r-i)},
\end{equation}
where
\begin{equation}
E^{\pm} = \left\{(\epsilon_1,\ldots,\epsilon_r)\in\{+,-\}^r : \prod_{i=1}^r \epsilon_i = \pm 1\right\}.
\end{equation}
For $\lambda = n\lambda_r$ or $\lambda = n\lambda_{r-1}$ all $g_i$ exponents in $\chi_{\lambda}$ are less than or equal to $n/2$, such that
only $\epsilon=(1,\ldots,1)$ contributes when the projection $P$ is taken. Evaluating the Vandermonde determinant we get
\begin{equation}
Pg^{-\rho}\Delta_r\chi_{\lambda}\chi_{n\lambda_r} = P g^{-\rho}\chi_{\lambda}\Delta_r^+ \prod_{i=1}^r g_i^{n/2}.
\end{equation}
For $\lambda = n\lambda_{r-1}$ this yields
\begin{equation}
Pg^{-\rho}\Delta_r\chi_{n\lambda_{r-1}}\chi_{n\lambda_r} = P g^{-\rho}\left(\Delta_r^- \right)^{-1} \mathcal{A}(0,r) \sum_{\epsilon\in E^-}
\prod_{i=1}^r g_i^{n/2 + \epsilon_i(n/2+r-i)}.
\end{equation}
Only the term with $\epsilon = (1,\ldots,1,-1)$ survives the projection. Thus
\begin{equation}
Pg^{-\rho}\Delta_r\chi_{n\lambda_{r-1}}\chi_{n\lambda_r} = P g^{-\rho}\left(\Delta_r^-\right)^{-1} \mathcal{A}(0,r)\prod_{i=1}^{r-1}
g_i^{(n+r-i)}.
\end{equation}
Since $\Delta_r^-=N_r^0$, the right hand side is equal to $Z_{r-1,n}$ as claimed. For $\lambda = n\lambda_r$ one obtains
\begin{equation}
Pg^{-\rho}\Delta_r\chi_{n\lambda_r}\chi_{n\lambda_r} = P g^{-\rho}\left(\Delta_r^-\right)^{-1} \mathcal{A}(0,r) \sum_{\epsilon\in E^+}
\prod_{i=1}^r g_i^{n/2 + \epsilon_i(n/2+r-i)}.
\end{equation}
Only the term with $\epsilon = (1,\ldots,1)$ survives the projection, so that
\begin{equation}
Pg^{-\rho}\Delta_r\left(\chi_{n\lambda_r}\right)^2 = P g^{-\rho}\left(\Delta_r^-\right)^{-1} \mathcal{A}(0,r) \prod_{i=1}^r g_i^{(n+r-i)} =
P_{\geq}Z_{r,n}.
\end{equation}
Since $P_0 Z_{r,n} = P Z_{r-2,n}$, we also have
\begin{equation}
P_0 g^{-\rho}\Delta_r \chi_{n\lambda_r}^2 = P Z_{r-2,n}.
\end{equation}
Since
\begin{equation}
P g^{-\rho}\Delta_r Z_{in} = \left(1 + \xi - P_0\right)P_{\geq}g^{-\rho}\Delta_r Z_{in},
\end{equation}
this confirms the Kirillov-Reshetikhin result for $z_{r-1,n}$ and $z_{rn}$. \qed
\end{proof}

\begin{remark}\label{D5}
As for $A_r$, the elements of $S(D_r,A_\ZZ)$ are parametrized by two group elements $g,h$. In particular one has $z_{1k}=N_k(g,h)/N_0(g)$,
where
\begin{equation}
N_k(g,h)=
\sum_{\epsilon\in\{-1,1\}^{r-1}}\sum_{\sigma\in {\mathcal S}^r} \mathrm{sgn}(\sigma)h_{\sigma(1)}^{\epsilon_1}
g_{\sigma(1)}^{\epsilon_1(r-1+k)}g_{\sigma(2)}^{\epsilon_2(r-2)}g_{\sigma(3)}^{\epsilon_3(r-3)}
g_{\sigma(r-1)}^{\epsilon_{r-1}}.
\end{equation}
\end{remark}

\begin{theorem}\label{D6}
Let $z$ be the image of an element of $S(D_r,A_n)$ in $(D_r,A_\NN)$. Then $z_{1m}=0$ for $m=n+2,n+3,\ldots,n+r$. Moreover
$\mathrm{d}z_{1,n+r}=0$ in $S(D_r,A_\ZZ)$ at the image of any element of $S(D_r,A_n)$ in this variety.
\end{theorem}

\begin{proof}
By assumption we have $z_{i,n+1}=1$ for all $i$. The $(D_r,A_\NN)$ equations immediately yield $z_{i,n+2}=0$ for all $i$. Now we put
$u_{im}=z_{im}$ for $i=1,\ldots,r-2$ and $u_{r-1,m}=z_{r-1,m}z_{rm}$, but leave $u_{im}$ undefined for $i=r$. Note that
$\mathrm{d}u_{r-1,n+2}=0$ at the image of any element of $S(D_r,A_n)$ in $S(D_r,A_\ZZ)$. To the extent that quantities are defined, the
$(D_r,A_n)$ equations for  the $u_{im}$ are invariant under an interchange of $r,n$. The discrete Hirota equations can be solved iteratively
with initial values $z_{1m}$. Accordingly, $u_{im}$ is the determinant of an $m\times m$ matrix with matrix elements $N_{jk}=x_{i+j-k}$,
where $x_j=z_{1j}$. We use induction on $m$. For $m=n+2$ we already have seen that the statement is true. Assume that $z_{1k}=0$ for
$k=n+2,\ldots,n+m$. Then consider the determinant which yields $z_{m,n+2}$. The minors with respect to the first row have vanishing
determinant, except for the last one, which is 1. Thus
\begin{equation}
u_{m,n+2}=(-)^{m+1}z_{1,n+m+1}.
\end{equation} \qed
\end{proof}

\begin{theorem}\label{D7}
\begin{equation*}
z_{1,m+2n+4r-2}=z_{1m}.
\end{equation*}
\end{theorem}

\begin{proof}
Evaluating the derivatives of $z_{1,n+r}$ with respect to the $h_i$ and imposing $\mathrm{d}z_{1,n+r}=0$ yields
\begin{equation}
g_i^{n+2r-1}=g_i^{-(n+2r-1)},
\end{equation}
for all $i$. This implies
\begin{equation}
z_{1,m+2n+4r-2}=z_{1m}.
\end{equation}
\qed
\end{proof}

\begin{remark}
As in the $(A_r,A_n)$ case one can show that this periodicity condition cannot be satisfied when the Weyl denominator vanishes, in other
words when there is a pair $i,j$ such that $g_i=g_j$ or $g_i=g_j^{-1}$. Indeed a calculation identical to the one for the $(A_r,A_n)$ case
yields a contribution to $z_{1n}$ proportional to $n^{k-1} g_i^n$, where $k$ is the number of vanishing factors in
$\sum_{j\not=i}(g_i-g_j)(g_i-g_j^{-1})$.
\end{remark}

We now sharpen the result $g_i^{2(n+2r-1)}=1$ obtained above to $g_i^{n+2r-1}=1$.

\begin{theorem}\label{D8}
\begin{equation*}
g_i^{n+2r-1}=1.
\end{equation*}
\end{theorem}

\begin{proof}
By the Weyl character formula we have $z_{1n}=N_n(g)/N_0(g)$, where
\begin{equation}
N_n(g)=\sum_{j=1}^r \left(g_j^{r-1+n}+g_j^{-(r-1+n)}\right)A_j,
\end{equation}
and
\begin{equation}
A_j=\sum_{\substack{\sigma\in{\mathcal S}^r\\\sigma(1)=j}}\sum_{\epsilon_2,\ldots,\epsilon_{r-1}\in\{1,-1\}}
\mathrm{sgn}(\sigma) g_{\sigma(2)}^{(r-2)\epsilon_2}\cdots g_{\sigma(r-1)}^{\epsilon_{r-1}}.
\end{equation}
Note that the Weyl denominator $N_0$ can be written as
\begin{equation}
N_0(g)=\sum_{\sigma\in{\mathcal S}^r}\sum_{\epsilon_1,\ldots,\epsilon_{r-1}\in\{1,-1\}}
\mathrm{sgn}(\sigma) g_{\sigma(1)}^{(r-1)\epsilon_1}\cdots g_{\sigma(r-1)}^{\epsilon_{r-1}},
\end{equation}
or
\begin{equation}
N_0(g)=\prod_{i=1}^r g_i^{-(r-i)}\prod_{i>j}(g_i-g_j)(g_i-g_j^{-1}).
\end{equation}
By the Weyl character formula, $g_i^{n+2r-1}=g_i^{-(n+2r-1)}$ for all $i$ implies
\begin{equation}
z_{1,n+r+k}=z_{1,n+r-k}.
\end{equation}
In particular, $z_{1,n+r+k}+z_{1,n+r-k}=0$ for $k=0,1,\ldots,r-2$ and $z_{1,n+r+k}+z_{1,n+r-k}=2$ for $k=r-1$.
This implies
\begin{equation}
\sum_{j=1}^r \left(g_j^{n+2r-1+k}+g_j^{-n-2r+1-k}+g_j^{n+2r-1-k}+g_j^{-n-2r+1+k}\right)A_j=0,
\end{equation}
for $k=0,\ldots,r-2$ and
\begin{eqnarray}
&&\sum_{j=1}^r \left(g_j^{n+2r-1+k}+g_j^{-n-2r+1-k}+g_j^{n+2r-1-k}+g_j^{-n-2r+1+k}\right)A_j \nonumber\\
&=&2\sum_{j=1}^r \left(g_i^{r+1} + g_i^{-r-1}\right)A_j.
\end{eqnarray}
Since
\begin{equation}
\sum_{k=1}^r\left(a_k^{r-j}+a_k^{j-r}\right)A_k=0,
\end{equation}
for $j=2,\ldots,r$, we can write
\begin{eqnarray}
&& \sum_{i=1}^r \left(g_i^{n+2r-1} + g_i^{-\left(n+2r-1\right)}\right)\left(g_i^k + g_i^{-k}\right) A_i \nonumber\\
&=& 2\sum_{i=1}^r \left(g_i^k + g_i^{-k}\right)A_i,
\end{eqnarray}
for $k=0,1,\ldots,r-1$.

By the Weyl character formula for $D_{r-1}$, all $A_k$ are different from zero if all $g_i$ are different. Moreover the matrix with entries
$a_k^{r-j}+a_k^{j-r}$, $j,k=1,\ldots,r$ has the same determinant as the Vandermonde matrix with entries $(g_k+g_k^{-1})^{r-j}$, which also
has non-vanishing determinant. Thus $g_i^{n+2r-1}-2+g_i^{-(n+2r-1)}=0$ for all $i$, which is equivalent to $g_i^{n+2r-1}=1$.

Conversely, the latter identities imply $z_{1,n+1}=1$ and $z_{1,n+k}=0$ for $k=2,\ldots,r$. This implies $z_{i,n+2}=0$ for $i=2,\ldots,r-2$
and $z_{i,n+1}=1$ for $i=2,\ldots,r-2$ by the $(D_r,A_n)$ equations. To fix the values of $z_{r-1,n+1}$ and $z_{r,n+1}$ one needs to choose a
preimage of $g$ in the connected cover of $SO(r,r)$, or equivalently a choice of squareroot of $g_1\cdots g_r$. \qed
\end{proof}

\begin{conjecture}\label{D9}
$S(D_r,A_\ZZ)$ is fibered over a maximal torus of $SO(r,r)$ modulo the action of the Weyl group.
\end{conjecture}

This statement seems to be much more difficult to prove than the analogous one for $S(A_r,A_\ZZ)$. Our current understanding is the following.
We have given a map from a dense subset of maximal torus pairs $(g,h)$ to $S(D_r,A_\ZZ)$. For any point $z$ in the latter space we can find
a sequence of pairs $g,h$ such that the image converges to $z$. We first show that the sequence can be restricted to a compact domain.
\begin{lemma}
For a sequence with limiting image in $S(D_r,A_\ZZ)$ the values of $|g_j|$, $|g_j|^{-1}$, $|h_j|$, $|h_j|^{-1}$ remain bounded
for all $j=1,\ldots,r$.
\end{lemma}

\begin{proof}
Consider a subsequence such that for a maximal subset $I$ of $\{1,\ldots,r\}$
the values of $|g_i|$, $|h_i|$ and their inverses remain bounded. Let $J$ be the complement of $I$. For $j\in J$ one can choose a subsequence
that either $|h_jg_j^n|\rightarrow\infty$ or $|h_jg_j^n|\rightarrow 0$ for sufficiently large $n$. In the first case put $\epsilon_j=1$,
in the second $\epsilon_j=-1$.
The character $z_{1n}$ can be written as
\begin{equation}
z_{1n} = \frac{\sum_{j=1}^r\left(h_jg_j^{n+r-1} + h_j^{-1}g_j^{-n-r+1}\right)}
{\prod_{k\neq j}\left(g_j-g_k\right)\left(1-g_j^{-1}g_k^{-1}\right)}.
\end{equation}
We relabel the indices so that $J=\{1,\ldots,k\}$. The character $z_{1n}$ can be written as
\begin{equation}
z_{1n} = \frac{\sum_{j=1}^r\left(h_jg_j^{n+r-1} + h_j^{-1}g_j^{-n-r+1}\right)}
{\prod_{k\neq j}\left(g_j-g_k\right)\left(1-g_j^{-1}g_k^{-1}\right)}.
\end{equation}
For $k>0$ consider the character $z_{kn}$, expressed as a determinant, as described in the
proof of Theorem 18. Looking only at the dominating term gives
\begin{eqnarray}
z_{kn} &=& \det\left(\frac{\sum_{j=1}^{|J|} h_jg_j^{n+r-1}}{\prod_{k\neq j}\left(g_j-g_k\right)\left(1-g_j^{-1}g_k^{-1}\right)}\right), \nonumber\\
&\sim & h_1^{\epsilon_1}\ldots h_k^{\epsilon_k}\left(g_1^{\epsilon_1}\ldots g_k^{\epsilon_k}\right)^n \times \mathcal{F},
\end{eqnarray}
where $\mathcal{F}$ denotes some non-zero factor independent of $n$. In the limit one finds
$|z_{kn}|\rightarrow\infty$, which is clearly impossible. \qed
\end{proof}

The previous lemma implies that any element $z\in S(D_r,A_\ZZ)$ can be obtained as an image of pairs $g,h$ where $g$ is fixed.
The remaining question is whether different choices of $g$ can lead to the same $z$.
As in the $A$-case one calculates for a generic element of $S(D_r,A_\ZZ)$,
\begin{equation}
Z_{2r,k}=\prod_{i=1}^r (g_i-g_i^{-1})^2,
\end{equation}
and $Z_{2r+1,k}=0$ for all $k$. If none of the $g_i$ is equal to $+1$ or $-1$ this implies that the sequence $z_{1k}$ and its translates span
a $2r$-dimensional vector space. In particular, the dimension of this vector space is bounded above by $2r$ for all elements in
$S(D_r,A_\ZZ)$. Let $p(T)$ be the monic polynomial of minimal order which annihilates the sequence $z_{1k}$. Whenever the dimension is
exactly $2r$, the map to the roots of $p$ yields a fibration over a maximal torus of $SO(r,r)$ modulo the action of the Weyl group. More
generally, consider the factorisation $p(T)=(T-1)^a(T+1)^bq(T)$, where $q$ is prime to $T^2-1$. Let $s$ be the order of $q$. When
$a+b+2+s=2r$, the solution belongs to the fiber corresponding to
\begin{equation}
g=\left(\underbrace{1,\ldots,1}_{a+1},\underbrace{-1,\ldots,-1}_{ab+1},g_{a+b+3},\ldots,g_{2r}\right),
\end{equation}
where $g_{a+b+3},\ldots,g_{2r}$ are the roots of $q$. It is possible, however, that $a+b+2+s<2r$. In this case different choices of $g,h$
may lead to the same sequence $z_{1k}$. The simplest example occurs for $a=b=0$, $s=2r-2$. Let $h_1=i, g_1=1$ and $g_k=i$ for $k=2,\ldots,r$,
with $h_2,\ldots,h_r$ chosen such that the sequence $z_{1k}$ remains finite. Then one obtains exactly the same sequence for $g'_1=-1$ and
$g'k=g_k$ for $k\neq 1$ and $h'k=-h_k$. The sequences $z_{2r-1,k}$ and $z_{2r,k}$ turn out to be different, however. We expect that this will
be true in all such situations.


\end{document}